\documentclass[conference]{IEEEtran}
\usepackage[latin9]{inputenc}
\usepackage{multirow}
\usepackage{amsmath}
\usepackage{amssymb}
\usepackage{graphicx}
\usepackage{esint}

\makeatletter

\DeclareTextSymbolDefault{\textquotedbl}{T1}
\providecommand{\tabularnewline}{\\}

\usepackage{multirow}

\DeclareTextSymbolDefault{\textquotedbl}{T1}
\providecommand{\tabularnewline}{\\}

\usepackage{array}
\usepackage{url}
\usepackage{multirow}
\usepackage{cuted}

\DeclareTextSymbolDefault{\textquotedbl}{T1}
\IEEEoverridecommandlockouts
\usepackage{cite}
\usepackage{amsfonts}
\usepackage{algorithmic}
\usepackage{textcomp}
\usepackage{xcolor}
\def\BibTeX{{\rm B\kern-.05em{\sc i\kern-.025em b}\kern-.08em
		T\kern-.1667em\lower.7ex\hbox{E}\kern-.125emX}}

\makeatother

\begin{document}
\title{An Initial Study of Human-Scale Blockage in sub-THz Radio Propagation
with Application to Indoor Passive Localization \thanks{Funded by the European Union. Views and opinions expressed are however
those of the author(s) only and do not necessarily reflect those of
the European Union or European Innovation Council and SMEs Executive
Agency (EISMEA). Neither the European Union nor the granting authority
can be held responsible for them. Grant Agreement No: 101099491 (project HOLDEN).\\
The measurement setup has been partially developed in the framework of SoBigData.it (Smart Radio Environment). SoBigData.it receives funding from European Union  \textemdash  NextGenerationEU-National Recovery and Resilience Plan (Piano Nazionale di Ripresa e Resilienza, PNRR) \textemdash Project: ``SoBigData.it \textemdash Strengthening the Italian RI for Social Mining and Big Data Analytics'' \textemdash Prot. IR0000013 \textemdash Avviso n. 3264 del 28/12/2021.}}
\author{\IEEEauthorblockN{F. Paonessa, G. Virone, S. Kianoush, A. Nordio, S. Savazzi}
\IEEEauthorblockA{\textit{\emph{Consiglio Nazionale delle Ricerche}}\emph{,} \textit{\emph{CNR-IEIIT}}, Italy.\\
 \linebreak{}
 }}
\maketitle
\begin{abstract}
This paper empirically investigates the body induced electromagnetic
(EM) effects, namely the human body blockage, by conducting indoor
measurement campaigns in the unexplored sub-THz W-band (75--110\,GHz)
and G-band (170--260\,GHz). The proposed analysis focuses on both the
alterations of channel frequency response induced by body presence,
fully or partially obstructing the line-of-sight (LoS) between transmitter and recevier, as well as
on the channel impulse response (CIR) for selected movements of the target, i.e. crossing the LoS of the radio link.
%Modelling of large scale parameters is also presented using a phantom body object.
The proposed study has applications in device-free radio localization and radio frequency (RF) sensing scenarios where the EM radiation or environmental
radio signals are collected and processed to detect and locate people
without requiring them to wear any electronic devices. Although preliminary,
the study reveals that discrimination of the blockage micro-movements
is possible, achieving higher precision compared to classical
RF sensing and localization using cm-scale wavelengths (2.4--6~GHz bands). 
\end{abstract}


\section{Introduction}

\label{sec:intro} Radio frequency (RF) sensing consists of an opportunistic set of
techniques capable to detect, locate, and track people in a monitored
area covered by ambient RF signals \cite{wilson-2010}.
Targeting the \emph{Integrated Sensing and Communication} paradigm~\cite{savazzi-1},
device free localization (DFL) systems can transform each RF node of the wireless network
covering the monitored area into a \emph{virtual sensor} capable of performing
sensing operations. RF sensing methods typically employ frequencies
in the unlicensed $2.4$--$5$~GHz ISM bands \cite{shit-2019}, and
up to $30$--$60$~GHz \cite{ojeda-2022}. In both cases RF signals
are characterized by wavelengths of only a few centimeters (cm-scale). However, multipath effects
strongly limit their accuracy, thus reducing their widespread acceptance.
On the contrary, sub-terahertz (sub-THz) radiation, generally characterized
by mm-scale wavelengths (mmWave), can be viewed as much better suited
for high-resolution body occupancy detection and vision applications
due to its very short wavelength and reduced multipath effects. The
diffusion of THz radios for 6G and beyond communication systems
\cite{thz} is thus expected to pave the way towards novel joint communication,
sensing and computing paradigms.

\begin{figure}
\centering \includegraphics[scale=0.45]{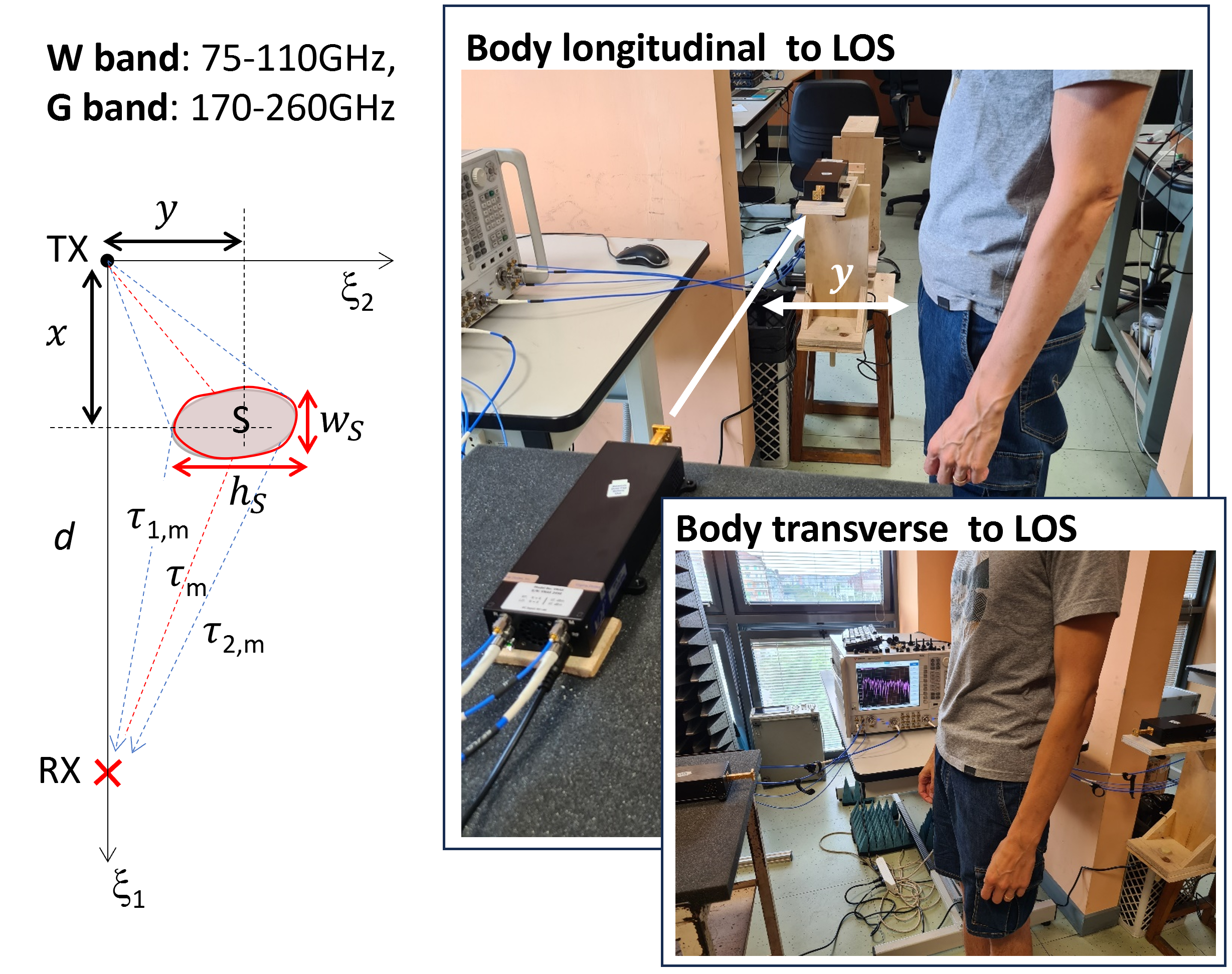} \protect\caption{\label{intro} (right) Measurement setup with Vector Network Analyzer (VNA), mmWave extenders (black boxes), and truncated waveguide antennas (gold-plated waveguide sections) in the laboratory environment. The white arrow identifies the line-of-sight. (left) 2D projection of the measurement setup: the shaded shape represents the human body. }
\end{figure}
Most of the sensing and vision techniques proposed for localization,
and behavior recognition \cite{CSI} such as Bayesian filtering, tomography
and holography \cite{kat,fall,holog}, require a detailed understanding
of the effects of human/object blockage and its physical/EM properties.
While body effects are well understood for classical ISM bands
\cite{ramp,scatt,mohamed-2017} and up to $30$--$60$\,GHz \cite{26G,bloc},
the human body blockage at frequencies beyond $100$\,GHz has been insufficiently
analyzed so far \cite{tap}. Although there are several studies proposing
physical and statistical models for assessing the impact of human
blockage on communication performance \cite{propagation_study,compcomm},
a limited body of knowledge exists regarding the EM and physical
characterization of body-induced fading effects. This includes characterization in terms of both time and frequency domains, which is critical for human-scale sensing applications.
%. in terms
%of time and frequency domain characterization which is critical in
%human-scale sensing applications.

In this paper, we provide a first study targeting the characterization
of human blockage in the unexplored W-band ($75$--$110$\,GHz) and
G-band ($170$--$260$\,GHz). To this aim, we carried out an indoor measurement
campaign collecting the target-induced RF footprints for different
offsets to the blocker crossing the LoS path. Small
movements of the target as traversing the link are represented here
by using a full-body phantom, which accurately simulates a 
%is accurate enough to replace a
human subject in the blockage study. The main contributions of the
proposed study are: 
\begin{itemize}
\item empirical measurements of human body blockage process in the indoor
environment considering the dual (full) bands of $75$--$110$\,GHz and
$170$--$260$\,GHz; 
\item frequency and time domain analysis of the blockage effects, considering
the channel frequency response and the impulse response (CIR) for
different offsets to the target crossing the LoS path; 
\item characterization of the area around the link that is most sensitive
to the blocker when approaching the LoS, for both the considered bands. 
\end{itemize}
The paper is organized as follows. Section \ref{sec:Frequency-domain-analysis}
focuses on frequency domain analysis considering the W and G bands
separately. The study reveals that simple statistical analysis
of the received power is not sufficient to reconstruct the blocker
location. Next, in Section \ref{sec:Channel-Impulse-Response} we
consider the more challenging G band and characterize the CIR as an optimized set of power/amplitude and delay features. Perturbations of CIR features induced by the blockage
are analyzed. Planned future experiments and the general approach
to CIR characterization are summarized in the Section \ref{sec:Conclusions}.

\begin{table*}[tp]
\protect\caption{\label{parameters} VNA settings and antenna specs.}
\vspace{0cm}

\begin{centering}
\begin{tabular}{|c|c|c||c|c|}
\hline 
\multicolumn{3}{|c||}{\textbf{Network Analyzer settings}} & \multicolumn{2}{c|}{\textbf{Antenna specs}}\tabularnewline
\hline 
\multirow{3}{*}{\textbf{W}} & \textbf{Start/Stop Freq.}  & 75/110~GHz  & \textbf{Type}  & Truncated Waveguide  \tabularnewline
\cline{2-5} 
 & \textbf{Freq. spacing}  & 35\,MHz  &  \textbf{FHPBW }  & about 90${^{\circ}}$ \tabularnewline
\cline{2-5} 
 & \textbf{Resolution BW}  & 1\,kHz  & \textbf{Polarization}  & Vertical \tabularnewline
\hline 
\multirow{3}{*}{\textbf{G}} & \textbf{Start/Stop Freq.}  & 170/260\,GHz  & \textbf{Antenna gain}  & 6-8\,dBi\tabularnewline
\cline{2-5} 
 & \textbf{Freq. spacing}  & 90\,MHz  & \textbf{TX power}  & 4-18\,dBm\tabularnewline
\cline{2-5} 
 & \textbf{Resolution BW}  & 1\,kHz  & \multicolumn{1}{c}{} & \multicolumn{1}{c}{}\tabularnewline
\cline{1-3} 
\end{tabular}
\par\end{centering}
\medskip{}
 \vspace{-0.6cm}
\end{table*}

\section{Measurement setup and environment}

The measurement sessions were conducted %took place 
in an indoor laboratory environment
with an approximate size of $4\times4$~m and a floor-to-ceiling height of $3$\,m. 
To describe the system geometry we considered the 2D Cartesian coordinate system depicted in Fig.~\ref{intro} whose origin coincides with the transmitter (TX). This coordinate system identifies an horizontal plane placed at about $h=1$\,m from the floor.
As shown in the figure, TX and RX nodes are spaced by $d=0.92$\,m apart, along the $\zeta_1$ axis. The target obstructing the LoS link is represented by
a simplified phantom (metal cylinder with  height 50\,cm and diameter 6\,cm) which is used to replace the human body. The projection of the target baricenter on the coordinate system has coordinates $(x,y)$.
The complex frequency domain channel coefficient~\cite{virone2007} were measured using a vector network analyzer (VNA) connected to a pair of mmWave extenders for both W and G bands. For both bands
we conducted 6 experiments, by varying the 
target offset w.r.t. the LoS path connecting the TX and RX, namely we considered $y \in\{0,3,6,12,25,50\}$\,cm.
Note that, in the case $y=0$ the target is completely obstructing the LoS path.
For each band and target offset, the VNA collected the transmission coefficients, 
measured at equally spaced frequencies, as detailed in Tab.~\ref{parameters}.
%1) $y=y_{1}=0$~cm
%(target blocking the LoS path); 2) %$y=y_{2}=3$~cm, 3) $y=y_{3}=6$~cm, 4) %$y=y_{4}=12$~cm; 5) $y=y_{5}=25$~cm, 6) %$y=y_{6}=50$~cm.
Furthermore, we also measured the channel frequency response in the target's absence with the purpose of calibrating the system.

\begin{table}[tp]
\protect\caption{\label{av_st} Frequency domain analysis: body induced RF attenuation
average and standard deviation}
\vspace{0cm}

\begin{centering}
\begin{tabular}{|c|c|c|}
\cline{2-3} 
\multicolumn{1}{c|}{} & \textbf{W band (}$\overline{A}_{i}$\textbf{,$\triangle A_{i}$)}  & \multicolumn{1}{c|}{\textbf{G band (}$\overline{A}_{i}$\textbf{,$\triangle A_{i}$)}}\tabularnewline
\hline 
$y_{1}=0$~cm  & ($14.3$~dB, $5.3$~dB)  & ($16.6$~dB, $6.8$~dB)\tabularnewline
\hline 
$y_{2}=3$~cm  & n.a.  & ($2.67$~dB, $4.45$~dB)\tabularnewline
\hline 
$y_{3}=6$~cm  & ($0.36$~dB, $1.52$~dB)  & ($-0.1$~dB, $3.4$~dB)\tabularnewline
\hline 
$y_{4}=12$~cm  & ($-0.05$~dB, $1.03$~dB)  & ($-0.08$~dB, $3.18$~dB)\tabularnewline
\hline 
$y_{5}=25$~cm  & ($-0.08$~dB, $0.69$~dB)  & ($-0.06$~dB, $2.9$~dB)\tabularnewline
\hline 
$y_{6}=50$~cm  & ($.002$~dB, $0.42$~dB)  & \multicolumn{1}{c|}{($-0.05$~dB, $2.85$~dB)}\tabularnewline
\hline 
\end{tabular}
\par\end{centering}
\medskip{}
 \vspace{-0.6cm}
 
\end{table}
\begin{figure}
\centering \includegraphics[width=1.0\columnwidth]{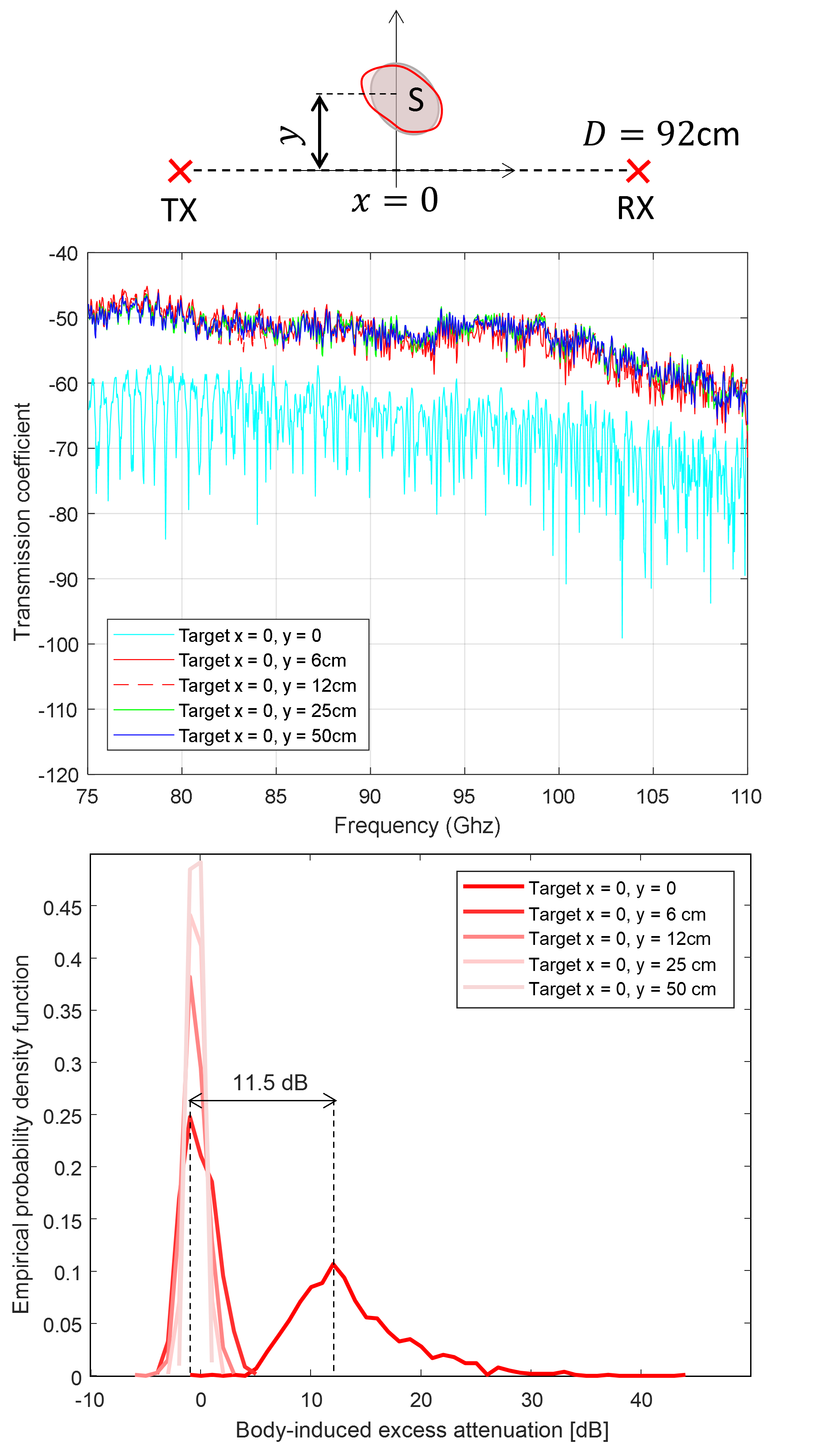} \protect\caption{\label{wband} W band: amplitude of transmission coefficient (top) and
body-induced excess attenuation probability functions
(bottom) plotted versus frequency for target positions $y \in \{0,3,6,12,25,50\}$\, cm.}
\end{figure}
\begin{figure}
\centering \includegraphics[width=1.0\columnwidth]{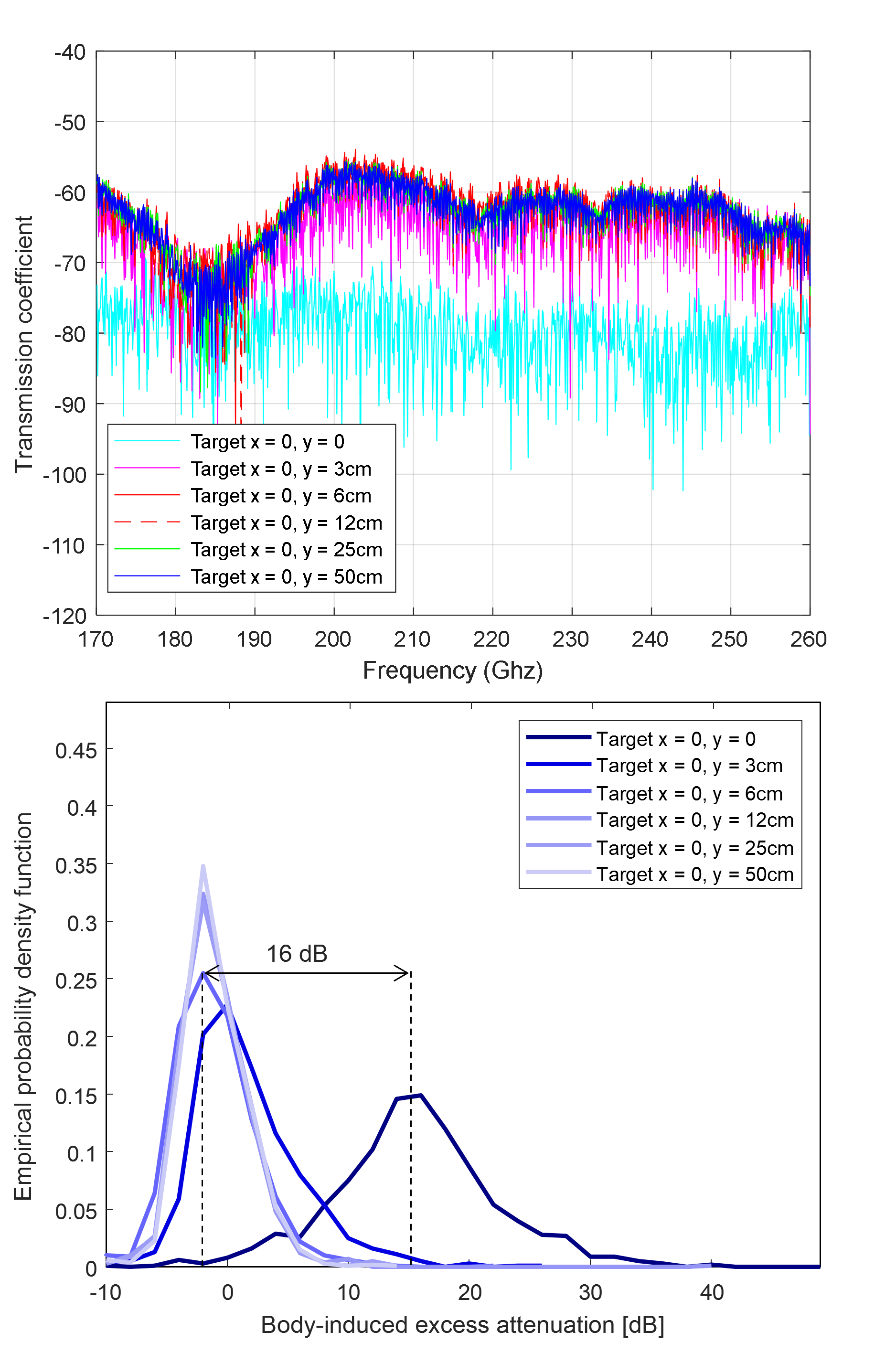} \protect\caption{\label{gband} G band: amplitude of transmission coefficient (top) and body-induced excess attenuation probability functions (bottom) plotted versus frequency, for target positions $y \in \{0,3,6,12,25,50\}$\,cm.}
\end{figure}

\begin{figure*}
\centering \includegraphics[scale=0.63]{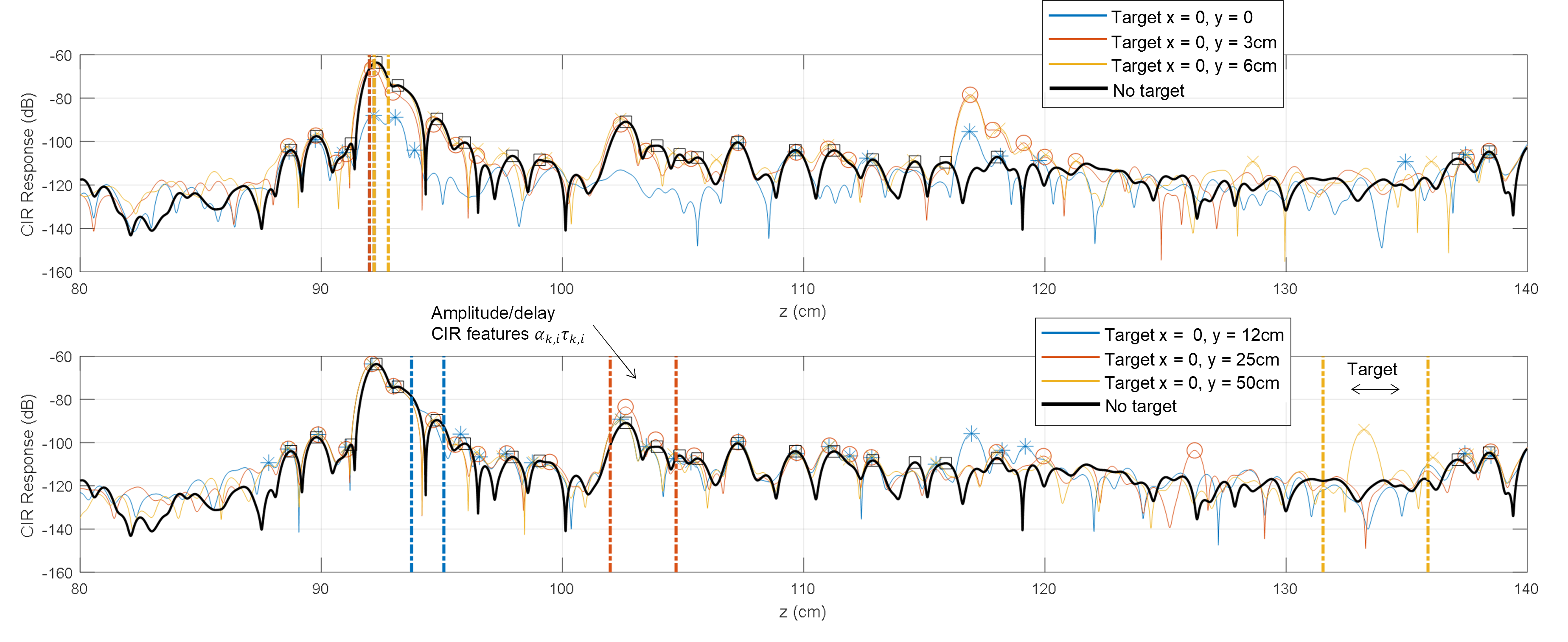} \protect\caption{\label{gband-1} Channel Impulse Response (dB) plotted versus the path length (cm) for target offsets
 $y \in \{0,3,6\}$ (top) and $y\in \{12,25,50\}$\,cm
(bottom). The CIR response in the absence of target is represented by the solid black lines. The main perturbations due to the target presence are highlighted.}
\end{figure*}

\section{Frequency-domain analysis\label{sec:Frequency-domain-analysis}}

 In this Section we focus on frequency domain analysis and we aim to verify the possibility
of discriminating the target position by monitoring the channel frequency response. In particular we discuss the problem of passive detection of the target considering 
the set of hypotheses $\{\mathrm{F}_{i}\}_{i=1}^6$, where the index $i$ is the identifier of the experiments previously described, and refers to the target offset $y=y_{i}$. The frequency responses in the W and G bands are evaluated separately. Specifically, each band was sampled at equally spaced frequencies $f_1, \ldots f_{N_f}$, where $N_f=1001$ is the number of sampling points.

During calibration, carried out with no target, we measured and stored the free-space transmission coefficient $T_0(f_k)$, for $k=1,\ldots,N_f$. 
We then denoted by $T(f_{k},i)$ the transmission coefficient measured during experiment $i$ at frequency $f_k$, with target offset $y_i$.
The RF attenuation due to the target at offset $y_i$ and measured at frequency $f_k$ is given by $A_{k,i}=-10\,\log_{10}\left[T(f_{k},i)\,/\,T_{0}(f_{k})\right]$.
The term $A_{k,i}$ can be considered a random variable with average
$\overline{A}_{i}=1/N_{f}\,\sum_{k=1}^{N_{f}}A_{k,i}$
and standard deviation $\triangle A_{i}=\sqrt{1/N_{f}\,\sum_{k=1}^{N_{f}}\left[A_{k,i}-\overline{A}_{i}\right]^{2}}$.
We also denote by $\Pr\left(A_{k,i}\,|\mathrm{F}_{i}\right)$ 
the sample distribution of $A_{k,i}$, conditioned to the hypothesis $\mathrm{F}_{i}$.
The detection of the target offset
can be obtained by one-vs-all classification using the log-likelihood-ratios (LLRs) for each hypotheses pair: 
\begin{equation}
\mathrm{\Gamma}_{i,j}=\mathrm{log}\left[\frac{\Pr\left(A_{k,i}\,|\mathrm{F}_{i}\right)}{\Pr\left(A_{k,j}\,|\mathrm{F}_{j}\right)}\right],\forall(i,j),i\neq j,\label{eq:llr}
\end{equation}
and applying a majority voting policy.

%In what follows we provide a frequency-domain analysis of the body-induced
%RF attenuations for both the considered bands. Data are organized
%according to the 6 body configurations previously described. 
In Figure~\ref{wband} and in Table \ref{av_st} we show the results obtained in the W band. Specifically, the table
reports the mean, $\overline{A}_{i}$, and the standard deviation $\triangle A_{i}$, for all the monitored
target positions whereas in the figure we plotted the amplitude of the transmission coefficient and the 
sample distribution $\Pr\left(A_{k,i}\,|\mathrm{F}_{i}\right)$.

We observe that, excluding the case $i=1$, where the target is completely blocking the LoS link, the sample distributions do not significantly differ, except for negligible variations of the standard deviation. This causes
the decision regions of the LLR tests~\eqref{eq:llr} for the  cases $i>1$ to be almost overlapped, with average separation less
than $1$~dB, thus penalizing the detection performance. Instead, the body induced average RF attenuation for the target blocking the LoS, i.e., $i=1$ and $y=y_1=0$, is  $\overline{A}_{0}=14.3$\,dB, which is consistent with conventional scalar diffraction theory models~\cite{ramp}. A similar situation is observed when analyzing the G band (see Figure~\ref{gband} and Table \ref{av_st}). Compared to the W band, much stronger variations of the RF attenuation, $\triangle A_{i}$ (up to 10\,dB) are observed, for target offset $y\le 3$\, cm. The body induced average RF attenuation for $y=0$\, cm is $\overline{A}_{0}=16.6$~dB.
Fron the figure, the only clearly recognizable target positions offsets
are $y=0$\,cm and $y=3$\,cm.

\section{Channel Impulse Response characterization and analysis\label{sec:Channel-Impulse-Response}}

In the following, we model the CIR as a
discrete set of power/amplitude and delay features. The presence of
the blockage introduces new multi-path components and may enhance or attenuate existing ones.
%as well as enhancing or attenuating existing ones. 
The target effects on propagation are
here modelled as alterations of selected multipath features. These
alterations can be further processed to reveal the unknown position
of the target with respect to the LoS link. 

The CIR features are obtained from the power delay profile (PDP) that is estimated from the complex measurements of the frequency response. PDPs are used in several
motion sensing and localization-based applications~\cite{xie-2018},
combined with Direction of Arrival estimation~\cite{wlan}. In what
follows, and according to previous analysis, we focus on the measurements in the G band. 

\subsection{CIR response}

The CIR response, $h_0(\tau)$, in absence of the target is represented
by a discrete dataset of $K_{0}$ amplitude $\mathbf{\boldsymbol{\alpha}}_{0}=[\alpha_{k,0}]_{k=1}^{K_{0}}$
and delay $\mathbf{\mathbf{\boldsymbol{\tau}}}_{0}=[\tau_{k,0}]$$_{k=1}^{K_{0}}$
features as
\begin{equation}
h_{0}(\tau|\mathbf{\boldsymbol{\alpha}}_{0},\mathbf{\mathbf{\boldsymbol{\tau}}}_{0})=\sum_{k=1}^{K_{0}}\alpha_{k,0}\delta(\tau-\tau_{k,0}),\label{eq:empty}
\end{equation}
where $K_{0}$ is the number of main observable multipath components, while $\alpha_{k,0}$
and $\tau_{k,0}$ are the amplitude and delay of path $k$, respectively. According to the  previously described blockage scenario, we assume that the target presence 
causes ``perturbations'' of the multipath components, in both amplitude and delay.
Such perturbations can be isolated from noisy observations of the PDPs so as to  estimate  the target offset $y_{i}$. 

When the target is located at position $y_i$, $i=1,\ldots,6$,  the CIR can be modelled as
\begin{equation}
h_{i}(\tau|\mathbf{\boldsymbol{\alpha}}_{i},\mathbf{\mathbf{\boldsymbol{\tau}}}_{i})=h_{0}(\tau|\mathbf{\widetilde{\boldsymbol{\alpha}}}_{0},\mathbf{\mathbf{\boldsymbol{\tau}}}_{0})+\sum_{k=1}^{\triangle K_{i}}\alpha_{k,i}\delta(\tau-\tau_{k,i}),\label{eq:cirpo}
\end{equation}
and be represented by an augmented dataset comprising  $K_{i}=K_{0}+\triangle K_{i}$ multipath components, whose amplitude and delays are 
\begin{equation}
\begin{aligned}\mathbf{\boldsymbol{\alpha}}_{i}={} & \left[\widetilde{\alpha}_{1,0},...,\widetilde{\alpha}_{K_{0},0},\alpha_{1,i},...,\alpha_{\triangle K_{i},i}\right],\\
\mathbf{\mathbf{\boldsymbol{\tau}}}_{i}={} & \left[\tau_{1,0},...,\tau_{K_{0},0},\tau_{1,i},...,\tau_{\triangle K_{i},i}\right].
\end{aligned}
\label{features}
\end{equation}
where $\widetilde{\alpha}_{k,0}=\rho_{k,i}\alpha_{k,0}$ and the parameter $\rho_{k,i}$ models the amplitude perturbation with respect
to that observed in the target's absence~\eqref{eq:empty}. The terms, $\alpha_{k,i}$ and $\tau_{k,i}$ model the \emph{ additional} multipath
components the target might give rise to.

\subsection{Analysis in the G band}
Figure \ref{gband-1} shows  the estimated PDP response, plotted versus the path length $z=c\tau$ (in cm), where $c$ is the speed of light for target offset $y \in \{0,3,6\}$\,cm (top) and $y= \{12,25,50\}$\,cm (bottom). The corresponding PDP measured without
the target is represented by solid black lines. 
By analyzing the results, we identify $K_{0}=6$ main multipath components in absence of the target.

However, the target presence can introduce up
to $3$ new components, i.e., $\triangle K_{i}\leq 3$.
The new CIR components also loosely depend on the target offset w.r.t. the LoS path, thus enabling the localization of the obstruction.

We observe that for $y\leq6$\,cm (i.e., $i=1,2,3$), no additional multipath components are introduced, w.r.t. $h_{0}(\tau)$. Therefore, for $i=1,2,3$, $\triangle K_i=0$. On the other hand, a significant attenuation
of the existing paths is observed, namely $\widetilde{\alpha}_{k,0}<\alpha_{k,0}$.
Excluding the first component, the CIR power features ($\widetilde{\alpha}_{k,0}$)
are subject to an attenuation $\rho_{k}$, $i=1,2,3$ ranging between $6$ and $21$\,dB. 

For larger offsets ($y>6$\,cm)
we observe that the target presence gives rise to new paths. In such case the value of $y$ can be inferred from the average path length of the new components. 

\section{Conclusions\label{sec:Conclusions}}
This paper provides an initial study of the characterization of human
blockage in the sub-THz band.
We conducted a measurement campaign focused on the collection of complex transmission coefficients for different offsets of the blocker crossing
the line-of-sight (LoS) path. Frequency and time domain analysis of
the blockage effects are considered. In particular, we analyzed the
channel frequency response and the impulse response (CIR) for different
blockage offsets.
Our experiments highlighted that the RF
attenuation due to the target presence are much stronger in the G band ($170$--$260$~GHz)
than in the W band ($75$--$110$~GHz). 
From the CIR response we also observed that the presence of the blockage gives rise to new multi-path components, while enhancing or attenuating the existing
ones. The target effects on the signal propagation are modelled as alterations 
of selected CIR multipath features. Importantly, the additional path components introduced by the target are characterized by a path length which depends on the target location. This allows 
to estimate the target offset w.r.t. the LoS link. These results are promising and are expected
to open new opportunities towards precise passive radio localization as well as targeting integrated sensing and communication applications.

\end{document}